\documentclass{sig-alternate}

\usepackage[utf8]{inputenc}
\usepackage{graphicx}
\usepackage{caption}
\usepackage{color}
\usepackage{bm}
\usepackage{xspace}
\usepackage{hyperref}
\newcommand\fnurl[2]{%
    \href{#2}{#1}\footnote{\url{#2}}%
}
\usepackage{balance}
\usepackage{subfig}

\definecolor{darkgreen}{RGB}{0, 128, 0}

\newcommand{\DatasetDescriptionWidth}{.30\linewidth}
\newcommand{\BiasFactorsWidth}{.32\linewidth}
\newcommand{\CorScatterWidth}{.32\linewidth}
\newcommand{\GiniWidth}{.9\linewidth}

\newcommand{\corUnifPrag}{$0.98$\xspace}
\newcommand{\corUnifLat}{$0.38$\xspace}
\newcommand{\corPragLat}{$0.47$\xspace}

\newcommand{\giniUnif}{$0.96$\xspace}
\newcommand{\giniPrag}{$0.83$\xspace}
\newcommand{\giniLat}{$0.7$\xspace}

\newcommand{\FigureDatasetDescription}{
\begin{figure*}[t!]
\centering

\subfloat[Number of Sessions]{\label{fig:dataset:numsessions}\includegraphics[width=\DatasetDescriptionWidth]{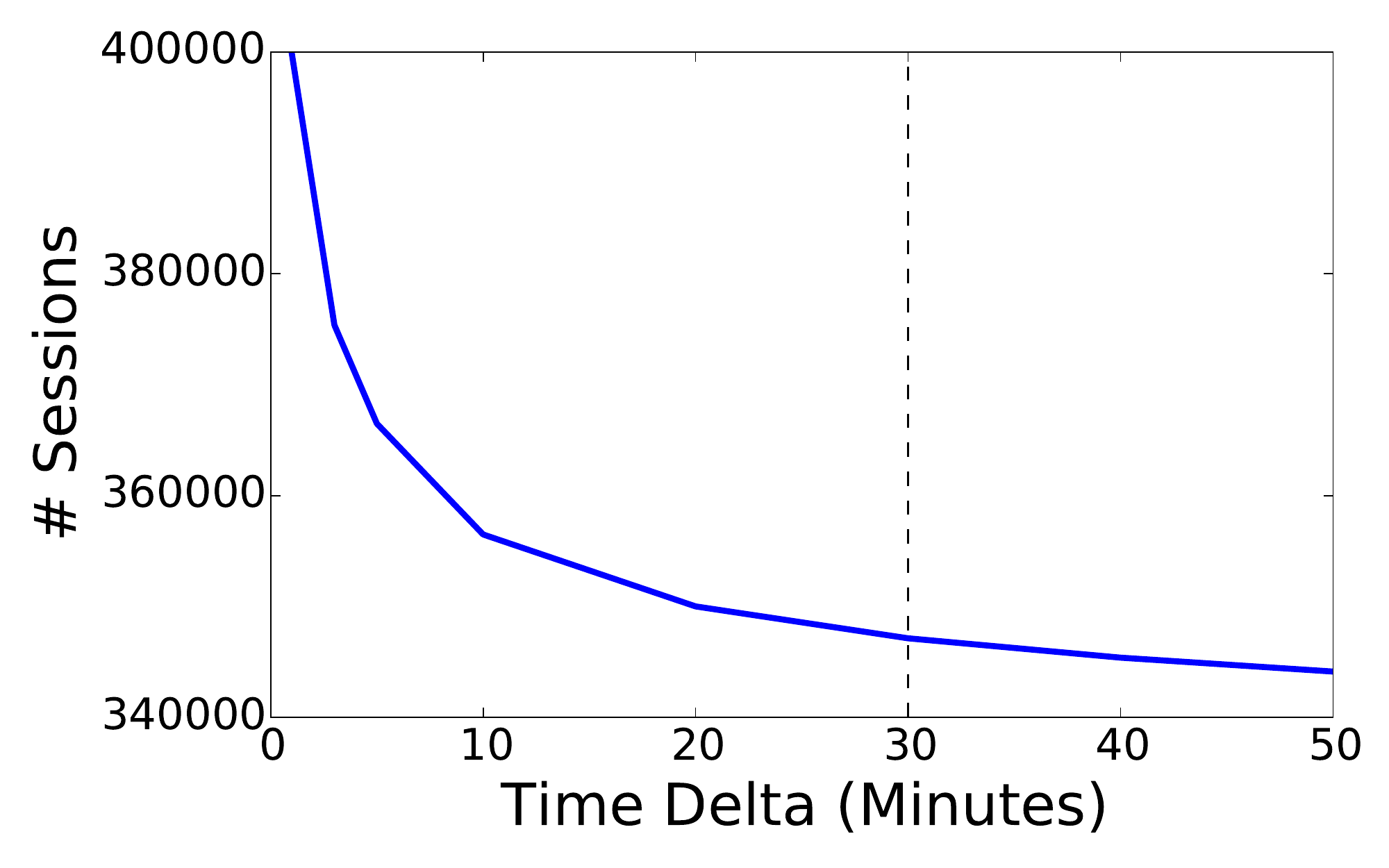}}
\hfill
\subfloat[Session Length]{\label{fig:dataset:avgsession}\includegraphics[width=\DatasetDescriptionWidth]{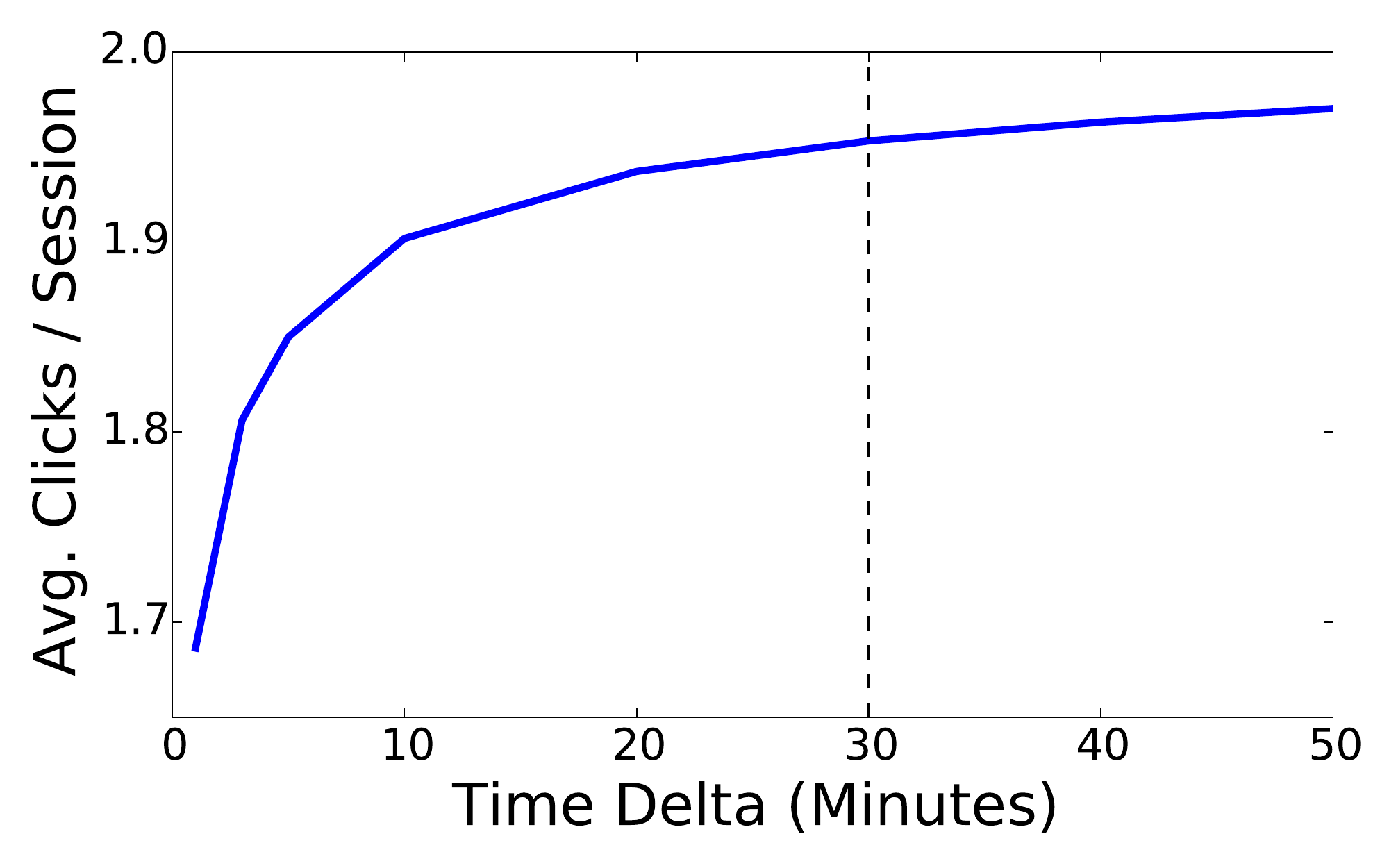}}\hfill
\subfloat[Session Distribution]{\label{fig:dataset:sessiondistr}\includegraphics[width=\DatasetDescriptionWidth]{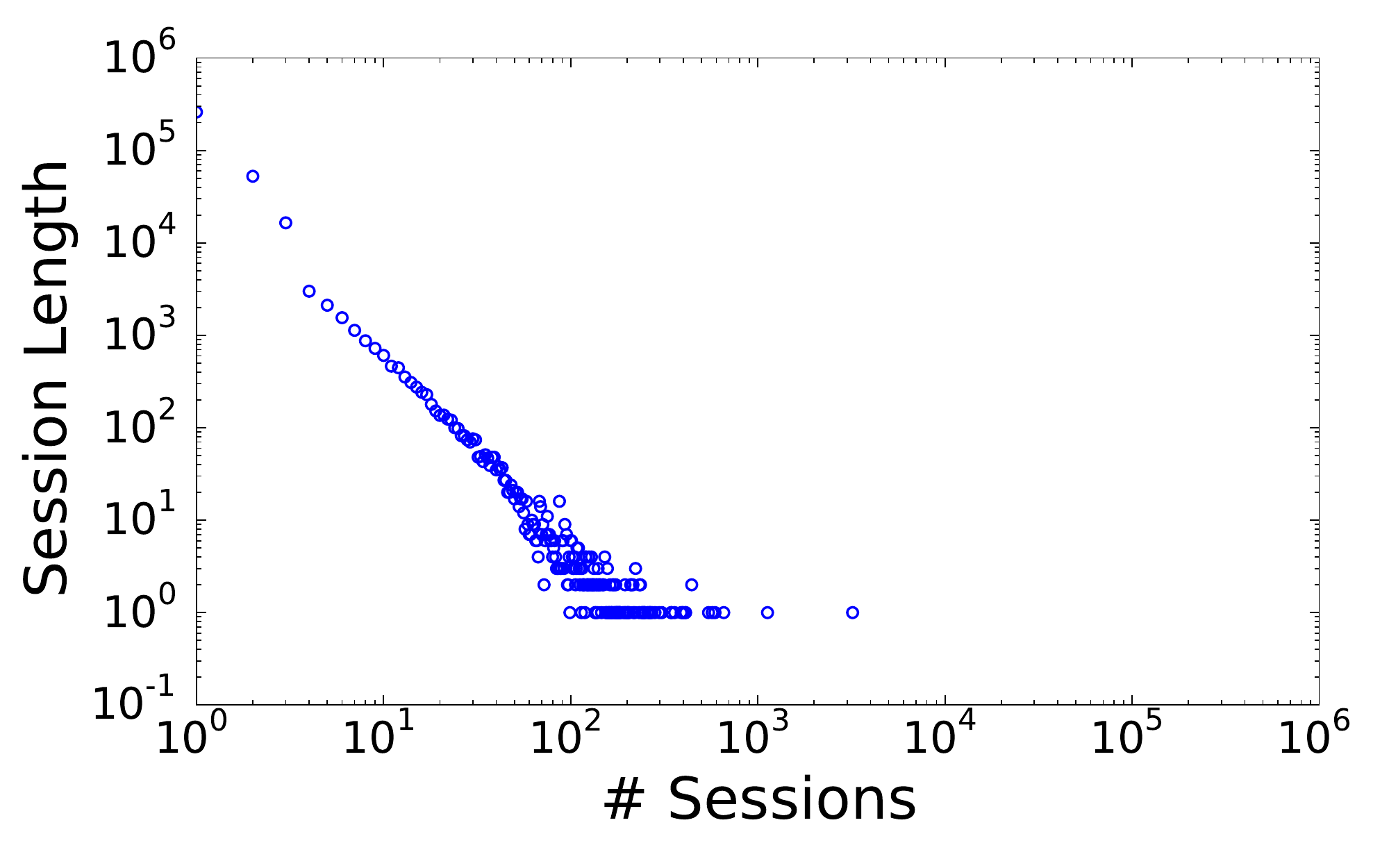}}
\vspace{-.6\baselineskip}
\caption{
\textbf{Dataset Description.} The figures depict characteristics of our dataset as well as the highly skewed heterogeneous distribution of the resulting sessions. The $y$-axis of Figure \ref{fig:dataset:numsessions} represents the number of sessions, while the $x$-axis represents the \emph{Time Delta}---the maximum time a user can spend between two clicks without creating a new session. We identified $30$ minutes to be a good compromise between numbers of sessions and session lengths. Figure (\ref{fig:dataset:avgsession}) depicts the \emph{average clicks} a user makes per session ($y$-axis) over different \emph{Time Deltas} ($x$-axis). We highlighted the chosen Time Delta of $30$ Minutes in both Figures (\ref{fig:dataset:numsessions} and \ref{fig:dataset:avgsession}). As can be seen, increasing the Time Delta would only result in a very small increase of session lengths. Figure \ref{fig:dataset:sessiondistr} visualizes the session lengths ($y$-axis) over the total number of observed sessions of specific length ($x$-axis). In our dataset we have many sessions of short lengths. With increasing session lengths, the number of observed sessions decreases, following a power-law distribution with $alpha=1.52$~\cite{powerlawfit}.
}
\label{fig:dataset}
\end{figure*}
}

\newcommand{\FigureHeatmap}{
\begin{figure*}[t!]
\centering
\subfloat[Uniform - Pragmatic]{\label{fig:heatmap:unif_pgra}\includegraphics[width=\CorScatterWidth]{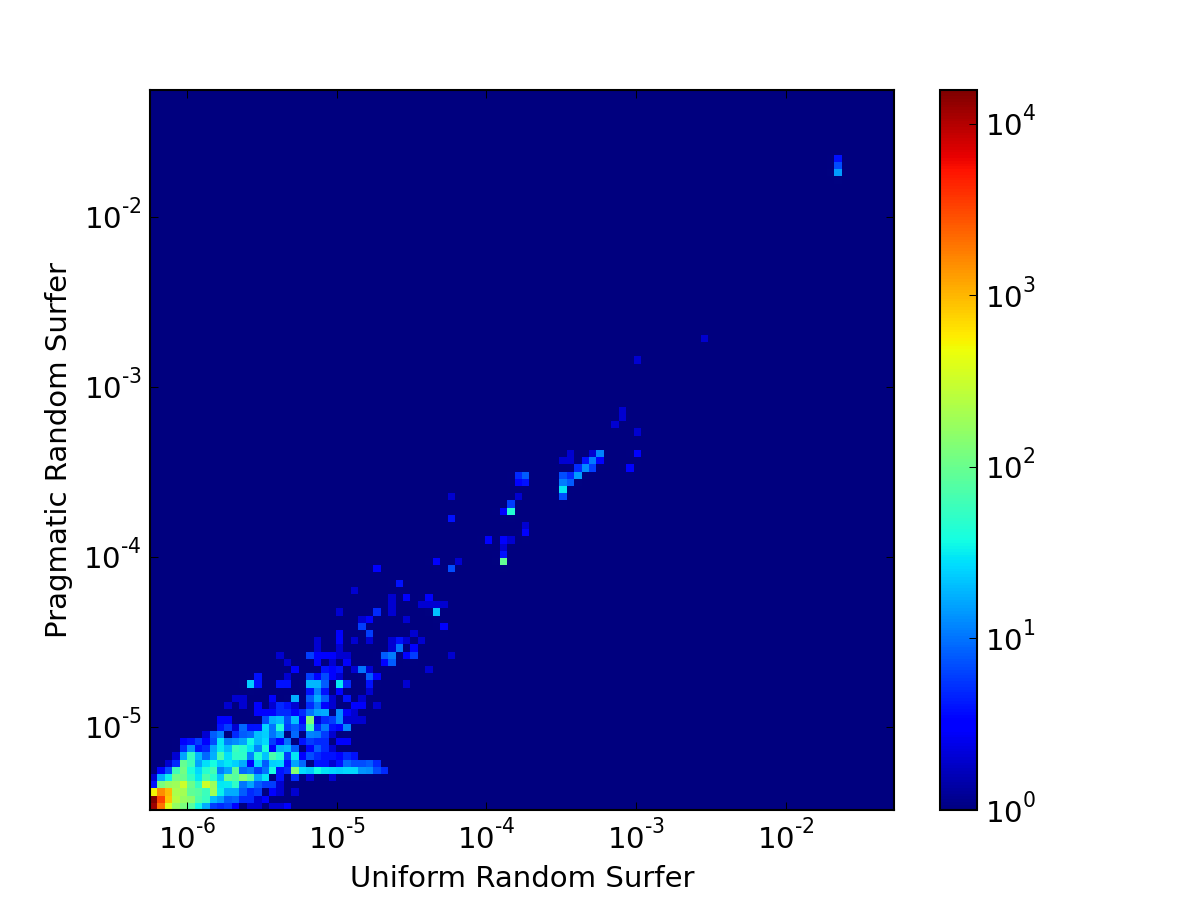}}
\subfloat[Uniform - Lateral]{\label{fig:heatmap:unif_pc}\includegraphics[width=\CorScatterWidth]{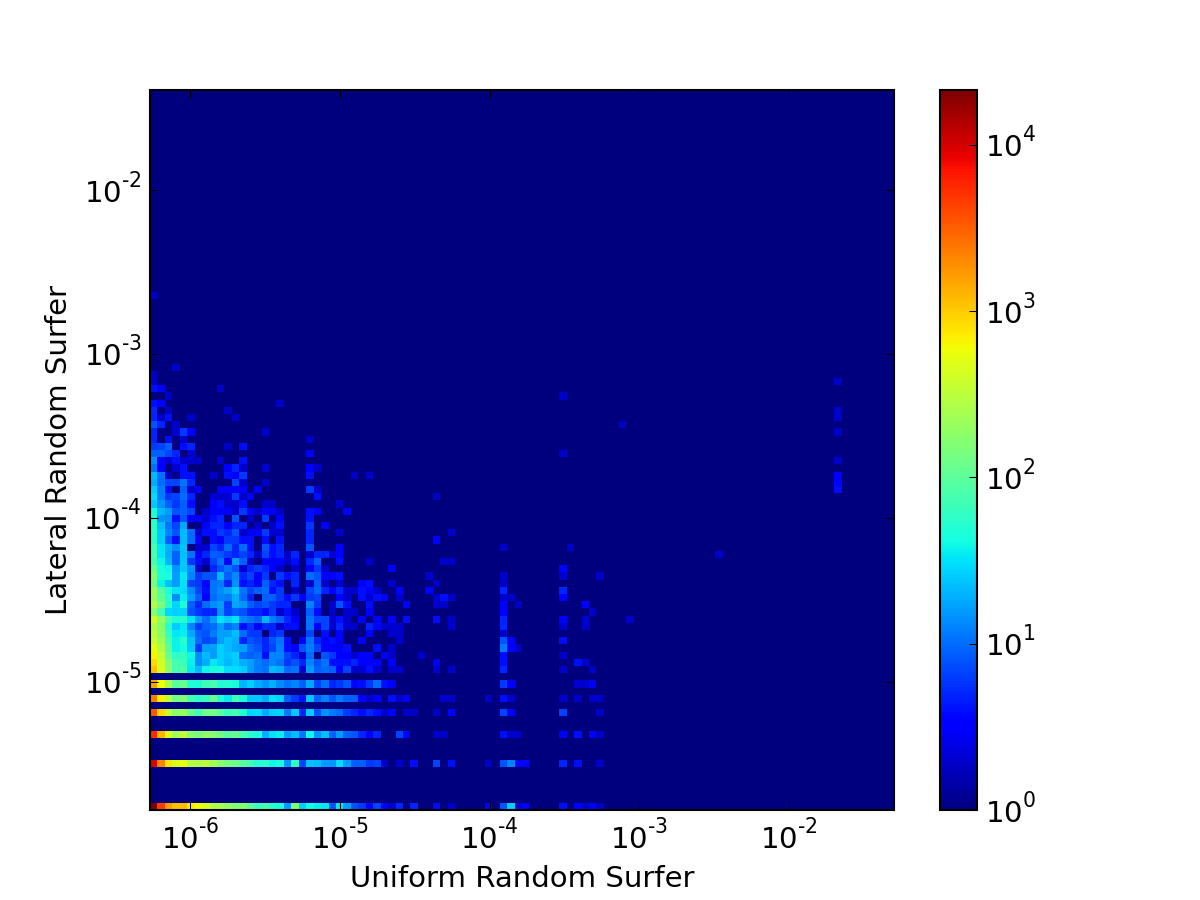}}
\subfloat[Pragmatic - Lateral]{\label{fig:heatmap:pgra_pc}\includegraphics[width=\CorScatterWidth]{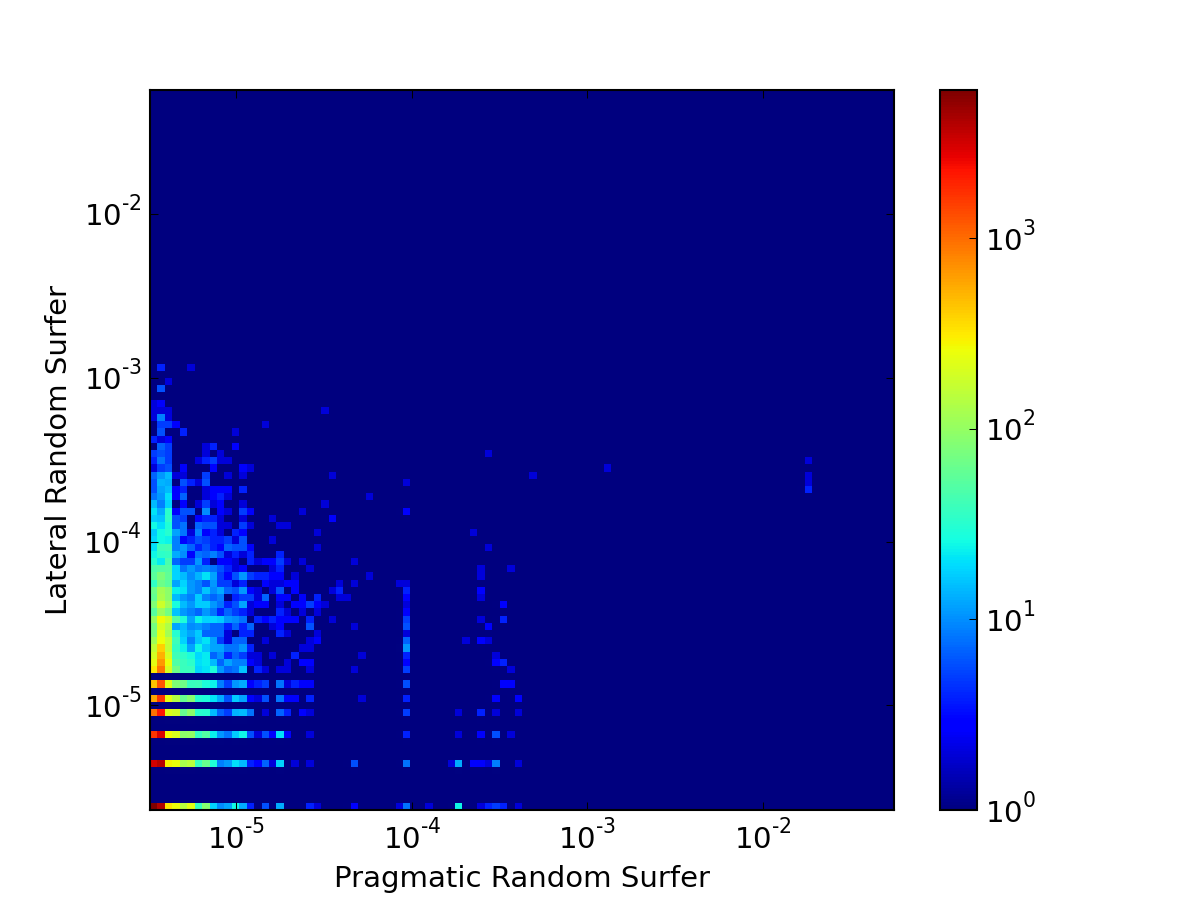}}
\caption{
\textbf{Correlation Scatter.} This figure depicts the correlation of the stationary distributions of all three random surfer models on a log-log scale. It shows binned elements of a scatter plot using a heat map. Colors refer to the amount of elements falling into a bin. Note that the color range is also on a log scale. We identified the strongest correlation between the uniform and pragmatic random surfer (Figure \ref{fig:heatmap:unif_pgra}) with a Pearson correlation coefficient of $\rho=$\corUnifPrag. In contrast, the correlation between the uniform and lateral random surfers (\ref{fig:heatmap:unif_pc}) is rather low with $\rho=$\corUnifLat. Figure \ref{fig:heatmap:pgra_pc} depicts the correlation of pragmatic and lateral random surfer with a Pearson correlation of $\rho=$\corPragLat. 
}
\label{fig:heat}
\end{figure*}
}

\newcommand{\FigureBiasFactors}{
\begin{figure*}[t!]
\centering

\subfloat[Uniform - Pragmatic]{\label{fig:biasfac:unif_prag}\includegraphics[width=\BiasFactorsWidth]{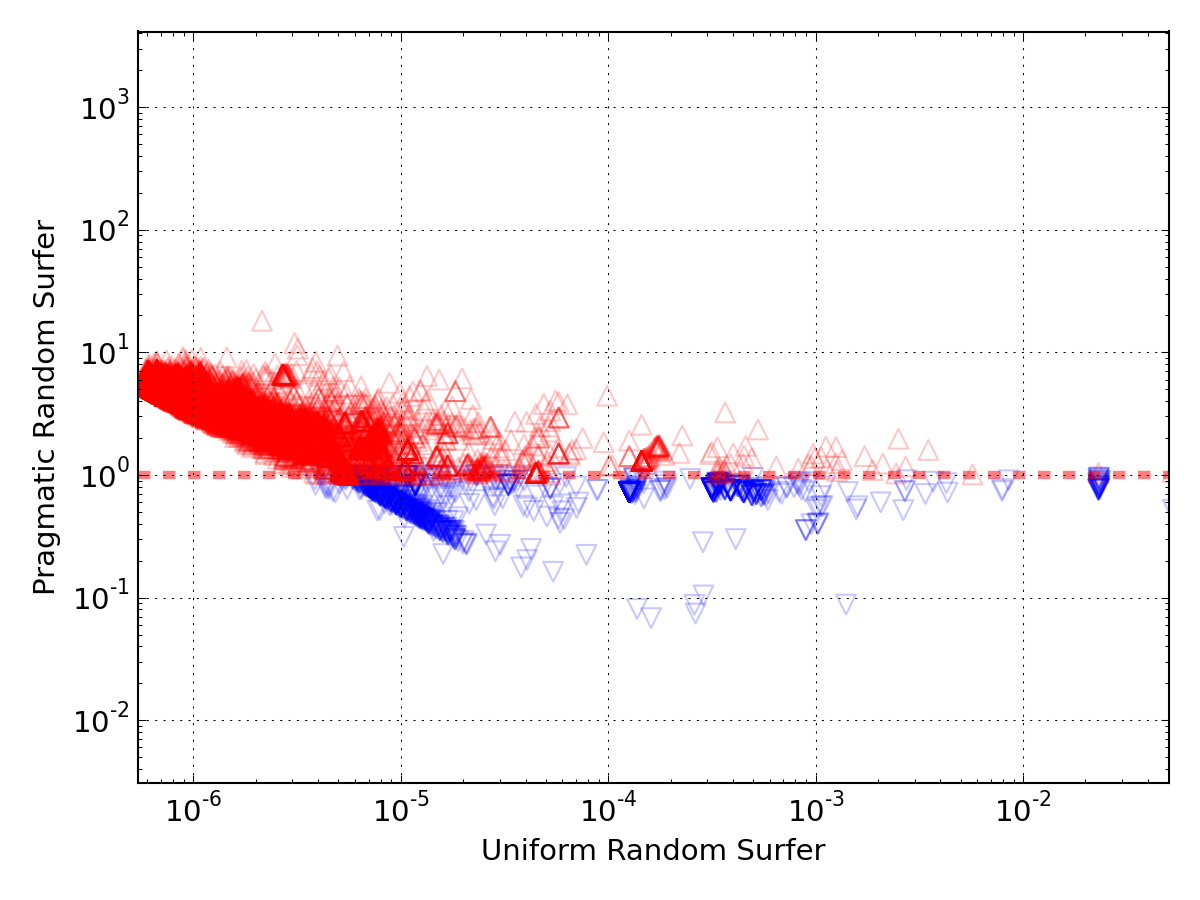}}
\hfill
\subfloat[Uniform - Lateral]{\label{fig:biasfac:unif_lat}\includegraphics[width=\BiasFactorsWidth]{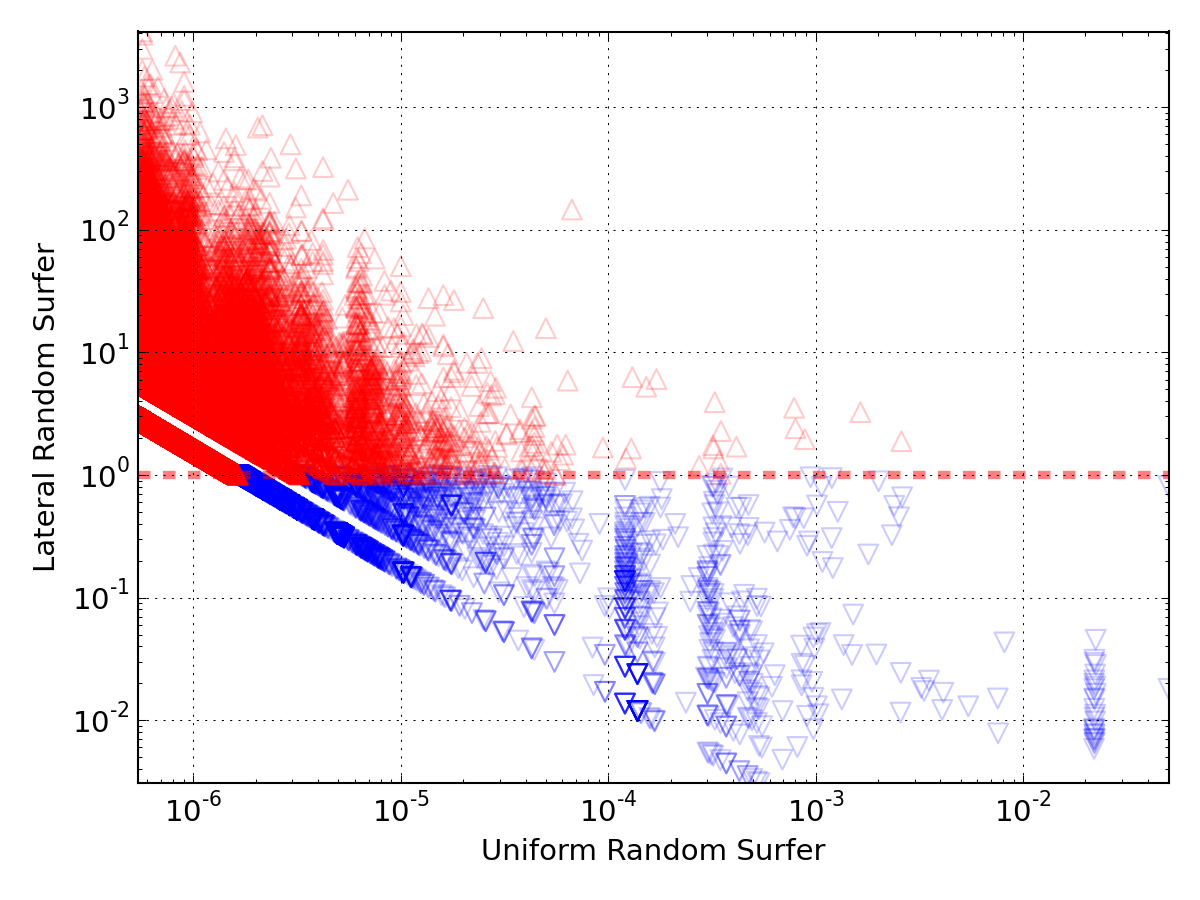}}
\hfill
\subfloat[Pragmatic - Lateral]{\label{fig:biasfac:prag_lat}\includegraphics[width=\BiasFactorsWidth]{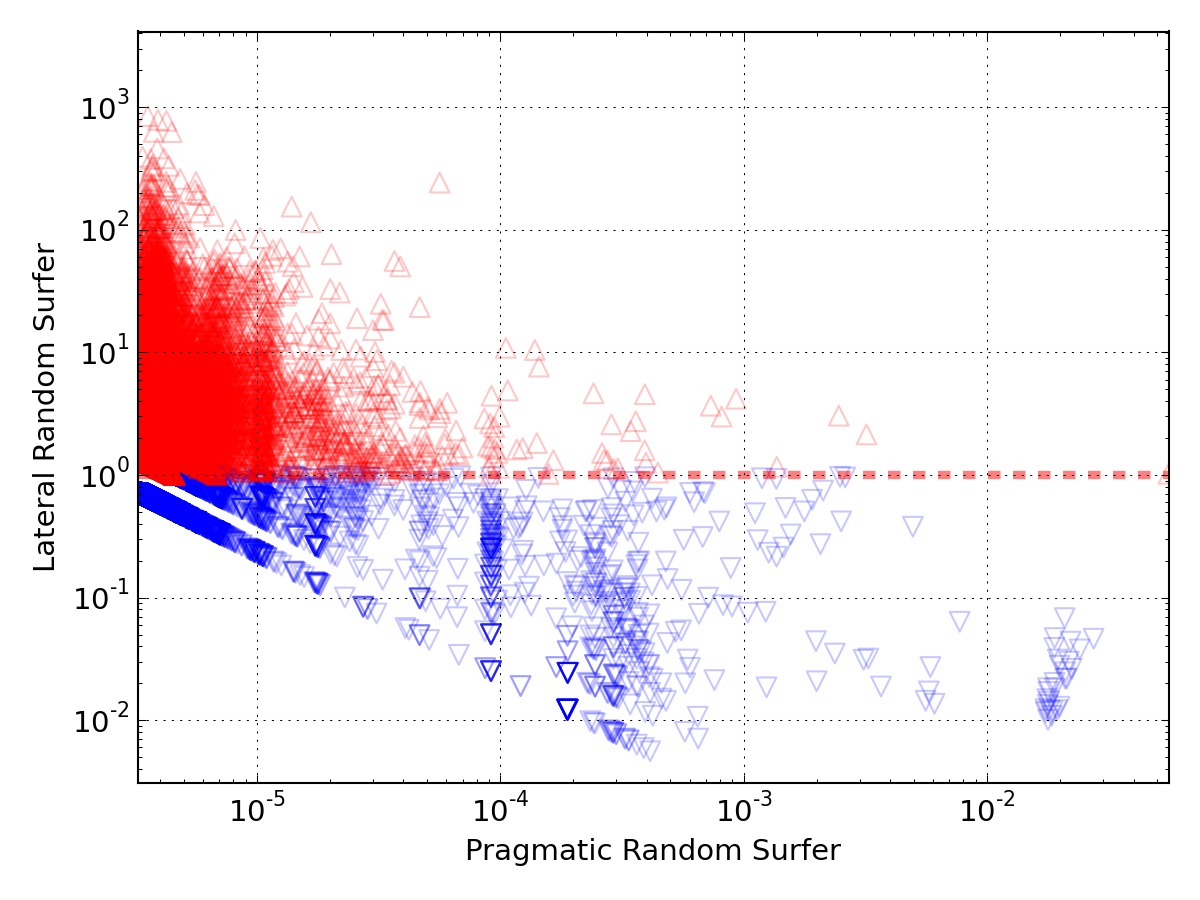}}

\includegraphics[width=0.4\linewidth]{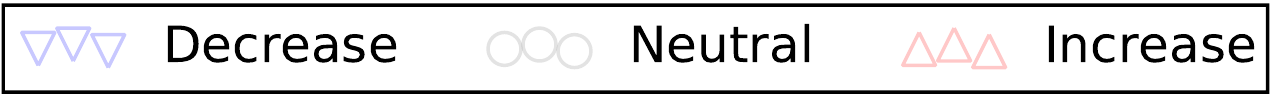}
\caption{
\textbf{Ratio of Stationary Probabilities.} The figures depict the ratio between stationary probabilities of pages for uniform, pragmatic and lateral random surfer. It contains basically the same information as Figure~\ref{fig:heat} transformed to ratios between values of the two stationary distribution under investigation. Figure~ \ref{fig:biasfac:unif_prag} shows the ratio between the uniform random surfer (as baseline) and pragmatic random surfer. Pages that are important for the uniform random surfer appear to be less important for the pragmatic random surfer. However, this difference is not significant (corroborated by a high correlation between those two random surfers), meaning that both surfers rate (nearly all of) the same pages as the most important ones. The ratio between the uniform random surfer and lateral random surfer (\ref{fig:biasfac:unif_lat}) shows that the latter strongly emphasizes pages with low stationary distribution values of the uniform random surfer. Thus, users have a higher tendency to visit just one page---nested deeper in the hierarchical network structure---of the Austria-Forum. Similar observations can be made for the pragmatic and lateral random surfers (\ref{fig:biasfac:prag_lat}). 
}
\label{fig:biasfac}
\end{figure*}
}

\newcommand{\FigureGini}{
\begin{figure}[t!]
\centering
\includegraphics[width=\linewidth]{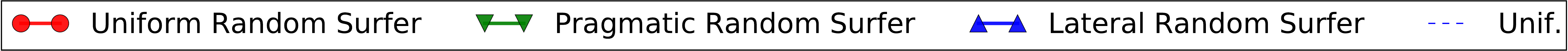}
\vspace{-0.8\baselineskip}

\subfloat[Gini]{\includegraphics[width=\GiniWidth]{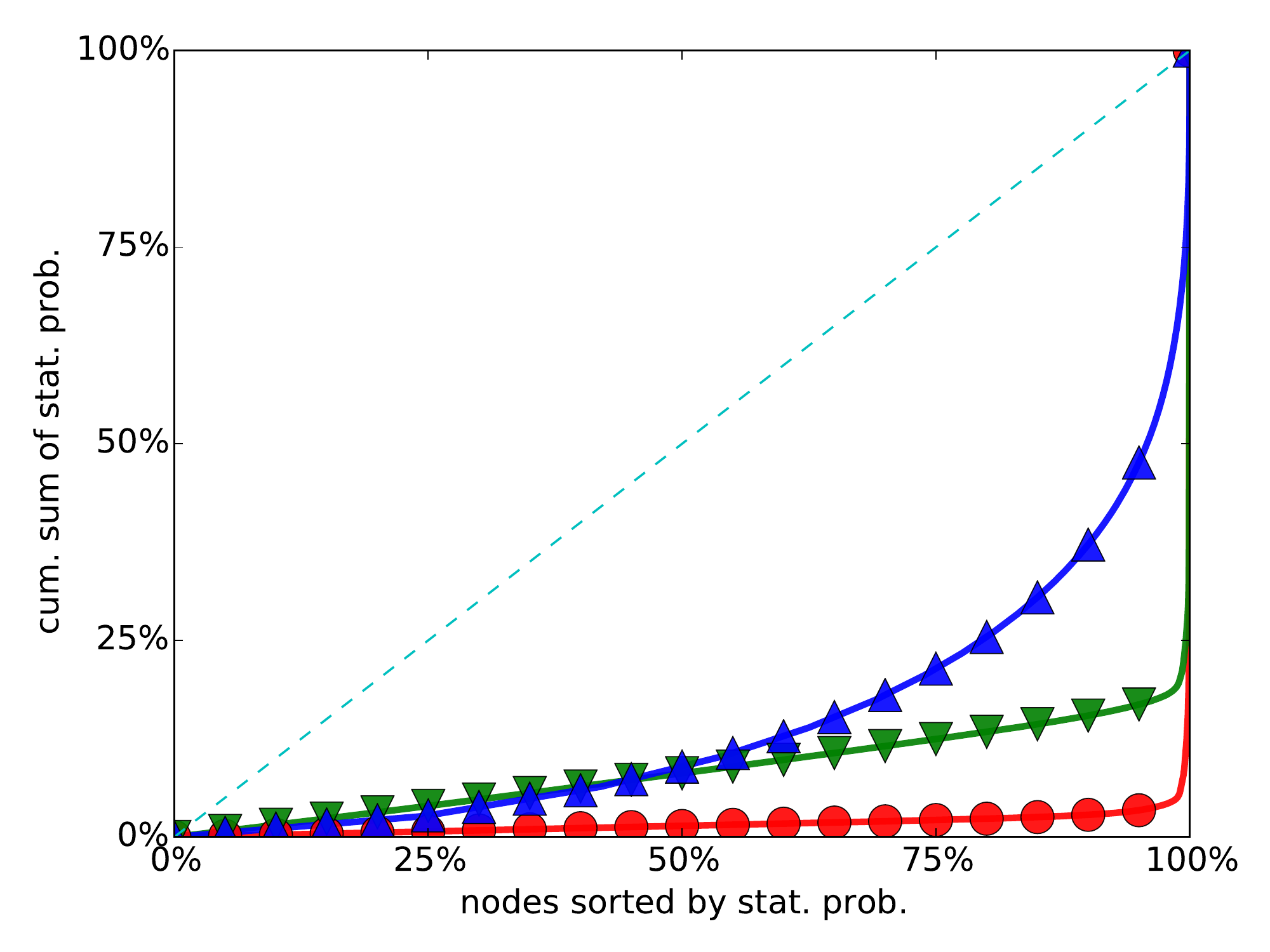}}
\caption{
\textbf{Lorenz-curves.} The plot depicts the Lorenz-curves of all three stationary distributions. We obtained the highest Gini coefficient of \giniUnif for the uniform random surfer, followed by the pragmatic random surfer with \giniPrag. The lateral random surfer achieved the lowest Gini coefficient (\giniLat). Thus, search engines (or other in-going links from external pages) likely point users to very specific pages of the Austria-Forum, tackling the problem of directing users to high importance pages, helping to mitigate the influence of popular websites on navigation behavior.}
\label{fig:gini}
\end{figure}
}
\newcommand{\DatasetHTTPRequestTable}{
\begin{table}[b!]
\caption{\textbf{HTTP-Request Log Entry.} The table shows the HTTP parameters which were logged and an example query entry where the user came from Google and visited the page of \textit{Waltraud Klasnic} which was successfully transmitted.}
\label{tab:dataset:httprequest}
\centering
\scalebox{.8}{
\begin{tabular}{r|l}
  \textbf{Date} & 2015-04-12 23:22:13,893\\
  \textbf{Method} & GET\\
  \textbf{Response Code} & $200$\\
  \textbf{Server Name} & austria-forum.org\\
  \textbf{Target} & [...]/Biographien/Klasnic,\_Waltraud\\
  \textbf{Request-Query} & None \\
  \textbf{Content-Type} & text/html;charset=UTF-8\\
  \textbf{Session-ID} & DC8F6B58BE968C906740853F4E6D4F41\\
  \textbf{Remote-IP} & 1.1.1.1 (for anonymity)\\
  \textbf{User-Name-Hash} & None \\
  \textbf{Referrer} & https://www.google.at/\\
  \textbf{User-Agent} & Mozilla/5.0 (iPad; CPU OS 8\_2 like Mac OS [...]\\
\end{tabular}
}
\end{table}
}

\newif\ifnight
\nighttrue
\nightfalse
\ifnight
    \usepackage{xcolor}
    \definecolor{bookColor}{cmyk}{0 , 0  , 0   , 0.2}  
    \color{bookColor}
    \usepackage{pagecolor,lipsum}
    \pagecolor{black!90}
\else
\fi

\begin{document}
\setcopyright{acmcopyright}
\conferenceinfo{i-KNOW '15,}{October 21-23, 2015, Graz, Austria}
\isbn{978-1-4503-3721-2/15/10}\acmPrice{\$15.00}
\doi{http://dx.doi.org/10.1145/2809563.2809598}
\title{Random Surfers on a Web Encyclopedia}

\numberofauthors{6} 

\author{
\alignauthor
Florian Geigl \\
       \affaddr{Graz University of Technology}\\
       \email{florian.geigl@tugraz.at}
\alignauthor
Daniel Lamprecht\\
       \affaddr{Graz University of Technology}\\
       \email{daniel.lamprecht@tugraz.at}       
\alignauthor Rainer Hofmann-Wellenhof\\
       \affaddr{Graz University of Technology}\\
       \email{rainer.hofmann-wellenhof@student.tugraz.at}
\and  
\alignauthor Simon Walk\\
       \affaddr{Graz University of Technology}\\
       \email{simon.walk@tugraz.at}
\alignauthor Markus Strohmaier\\
       \affaddr{GESIS and University of Koblenz-Landau}\\
       \email{strohmaier@uni-koblenz.de}
\alignauthor Denis Helic\\
       \affaddr{Graz University of Technology}\\
       \email{dhelic@tugraz.at}
}
\maketitle

\begin{abstract}
The random surfer model is a frequently used model for simulating user navigation behavior on the Web. Various algorithms, such as PageRank, are based on the assumption that the model represents a good approximation of users browsing a website. However, the way users browse the Web has been drastically altered over the last decade due to the rise of search engines. Hence, new adaptations for the established random surfer model might be required, which better capture and simulate this change in navigation behavior.
In this article we compare the classical uniform random surfer to empirical navigation and page access data in a Web Encyclopedia. Our high level contributions are (i) a comparison of stationary distributions of different types of the random surfer to quantify the similarities and differences between those models as well as (ii) new insights into the impact of search engines on traditional user navigation.
Our results suggest that the behavior of the random surfer is almost similar to those of users---as long as users do not use search engines. We also find that classical website navigation structures, such as navigation hierarchies or breadcrumbs, only exercise limited influence on user navigation anymore. Rather, a new kind of navigational tools (e.g., recommendation systems) might be needed to better reflect the changes in browsing behavior of existing users. 

\end{abstract}


\section{Introduction}
\label{sec:intro}

The last decades have seen immense growth of the Web, which now has an approximate size of over a billion \fnurl{Web pages}{http://www.internetlivestats.com/}.
The Web provides people around the world with access to a host of information resources and serves uncountable use cases, such as gathering information, studying, making financial transactions, shopping, or booking hotels. To find relevant information in this huge information system, Web users apply various information retrieval techniques. A very common---and probably the most basic and straight-forward---strategy consists of simply navigating between Web pages by traversing the provided hyperlinks from one Web page to another.
In many cases, users also jump directly to other Web pages by typing the \textit{URL} of the new target page in the browser address bar or by using a search engine and following one of the search results. These cases are typically referred to as \textit{teleportation}~\cite{brin}, as users ``teleport'' from the current Web page to another one.

The importance of Web navigation is even further amplified by an alternative informational retrieval strategy---Web search. Ranking algorithms used by search engines are based on variants of PageRank~\cite{brin}, which assigns weights based on hyperlinks. These ranking approaches assume a so-called random surfer~\cite{brin}---a model of a user who traverses the Web by following hyperlinks uniformly at random with a small chance of teleporting at each navigation step. 
In their original paper, Page and Brin~\cite{brin} suggested a damping factor of $0.85$, meaning that, for each step, users traverse hyperlinks with a probability of $85\%$, while exhibiting a probability of $15\%$ of teleporting to a page selected uniformly at random. The number of visits of an indefinitely navigating random surfer to each particular page is then a direct measure of page importance for Web navigation and is used to rank search results.

\noindent\textbf{Problem.} Although the random surfer model has proven to be extremely useful in practice, only a few studies have analyzed the capabilities of this model to imitate real user behavior in different contexts. Moreover, most of these studies concentrated on empirically analyzing the damping or teleportation factor (such as~\cite{Gleich2010}). In this work, we compare \emph{clickstream data of real users} with the \emph{random surfer model}. In particular, we are interested in analyzing how real users assess the importance of Web pages for navigation and how that assessment compares to that of the random surfer. 
Moreover, we also study to what extent the navigation of human users is influenced 
by the modern search engines. To this end, we analyze page view counts, which also account for landing pages from search engines.

In particular we are interested in answering the following research questions:
\begin{description}
\item[\textbf{RQ1 Comparison of a random surfer with real users.}] To what extent does a random surfer with teleportation imitate user navigation behavior?
\item[\textbf{RQ2 Influence of search engines.}] How do search engines affect how users access and navigate websites?
\end{description}

\noindent\textbf{Approach \& methods.} For our analysis, we first calculate the stationary distributions of a \textit{uniform} random surfer, traversing the information network uniformly at random with a teleportation probability of $15\%$. We then compare this stationary distribution with the stationary distribution of a \textit{pragmatic} random surfer, who selects the links with a probability that is proportional to the transition counts from empirical data (human users). For the pragmatic random surfer we again use $15\%$ teleportation probability.
Finally, we compare stationary distributions of both uniform and pragmatic random surfer with the stationary distribution (normalized page view count distribution) of a \textit{lateral} random surfer, which accounts for the lateral access from a search engine to a given website.

For the distribution comparison we calculate linear correlation factors and Gini coefficients to investigate the alignment of distributions, and the distributions' inequality, respectively.

\noindent\textbf{Contributions.}
Our high-level contribution is a better understanding of human navigation behavior and how it compares to a navigational model such as the random surfer model. 

Methodologically, we compute and analyze stationary distributions using a set of standard measures with a clear interpretation in the context of Web navigation. 

Empirically, we provide evidence that, despite its simplicity, a random surfer model is a very accurate model of basic human navigation behavior in our dataset. Our results suggest that the general navigation behavior of users is very much in line with the random surfer model---both assess the navigational page importance in a similar and highly skewed way, meaning that just a few pages are extremely important. These results also hold for cases where website operators decide to provide specific navigational structures (as in our dataset) such as navigational hierarchies. Users, as well as the random surfer, do not make any particular distinction between different types of links present on the website. However, the lateral access from search engines reduces the imbalances, at least for human users, and need therefore to be taken into account when modeling user navigational behavior.

\section{Related Work}
\label{sec:rel}

Our work relies heavily on the random surfer model, which is a simple but well-studied model for modelling navigation on the Web \cite{lovasz1993random, woess1994random}. Apart from navigation, the random surfer model has also been applied to a variety of different problems such as graph generation and graph analysis. In particular, Blume et al. \cite{blum2006random} used the model for the creation of web-graphs while \cite{randomwalkcom, rosvall2008maps,  zlatic2010topologically} have applied the model to detect community structures in networks.

Algorithms such as PageRank~\cite{brin, PRordertheweb} or HITS \cite{kleinberg1999authoritative}, use the random surfer as the basis for calculating node centralities in networks. PageRank includes a parameter to define the probability of teleportation for the random surfer. This parameter is often referred to as the \textit{damping-factor} $\alpha$, representing the probability that the random surfer traverses one of the links pointing away from the current node. With probability $1-\alpha$  it jumps to a network node chosen uniformly at randomly and continues surfing from there. In 2010, researchers have empirically measured this factor by analyzing clicktrails of humans and reported an estimated damping factor between $0.6$ and $0.72$ for the entire web \cite{Gleich2010}. In contrast, the damping factor for Wikipedia has been determined to be between $0.33$ and $0.43$. This difference in damping factors might be caused by the way users access Wikipedia---they use search engines that point them directly to the article of interest, rendering additional navigational efforts unnecessary. Researchers additionally investigated the connection between the damping factor and the convergence rate of the PageRank algorithm and found that it converges very fast for a value of $0.85$~\cite{haveliwala2003second, kamvar2003extrapolation}. However, in this paper we investigate the influence of the damping factor onto the stationary distribution of the random surfers.

\cite{qiu2006automatic} presented a framework that was able to personalize PageRank on a very small set of user-based clickdata for websites. Additionally, Al-Saffar and Heileman \cite{boundsoftopicsenspagerank} compared these personalized and topic-sensitive PageRank results with results from the unbiased (original) PageRank and came to the conclusion, that both ways of personalizing the PageRank produce a considerable level of overlap in the top results. 
In particular, the authors conclude that biases, which do not rely on the underlying link structure of the network under investigation, are needed to further improve the personalization of PageRank. In this paper we are interested in the stationary distribution of PageRank personalized by observed user transitions.

Researchers also looked closely into modeling human navigation behavior, using this biased random surfer model. For example, West and Leskovec \cite{west2012human} investigated human click trails of a navigation game played by humans on Wikipedia. Participants were asked to navigate from a given start article in Wikipedia to a specific target article, using as few clicks as possible. Using the results of this study, West and Leskovec \cite{west2012automatic} designed different features for steering a probabilistic random surfer. They also compared paths produced by the biased random surfer with those of humans and found that navigation of humans was based mostly on popularity and similarity biases. 
In 2013, Helic et al. \cite{Helic2013} compared click-trail characteristics of stochastically biased random surfers with those of humans. They concluded that biased random surfers can serve as valid models of human navigation. Furthermore, Singer et al. \cite{singer2014detecting} conducted experiments to find out whether human navigation is Markovian, meaning that the next click of a user is only dependent on the most recent click. They showed that on a page level, human navigation can be best explained by first-order Markov chains. This finding is particularly relevant for us, as it allows us to use simple biases which do not consider previously visited nodes of the random surfer for our experiments.

\section{Materials \& Methods}
\label{sec:meth}
\subsection{Datasets}
\label{subsec:datasets}
\noindent\textbf{Austria-Forum.} In this paper we use change and click data from Austria-Forum\footnote{\url{http://www.austria-forum.org}}, an Austrian web encyclopedia which was initially created more than two decades ago and restructured in 2009. Austria-Forum tries do distinguish itself from other well established web encyclopedias by providing mechanisms to counteract some specific drawbacks: For instance, Austria-Forum tries to fight against the apparent (personal) biases of anonymous contributions by having (and enforcing) approved and named authors as the only contributors to the knowledge base. Authors are mostly academics well-established  in their field, which has the positive aspect of thoroughness since they exhibit a personal interest not to produce literature of low quality. 
As the name suggests, the information published is geographically limited to all things concerning the country of Austria. Compared to other resources on the web, Austria-Forum tries to transmit the knowledge on a more granular level. Not only does it provide users with several differently scoped articles, but also with entire digitized books as Web Books on a variety of different cultural and historical aspects of Austria.
In order to increase the amount of displayed content, Austria-Forum added the capability of including entire pages from different external domains into their Wiki (e.g., of the German Wikipedia). 

Most of the interactions of a user with an encyclopedia are limited to single page views, usually generated by direct requests via a search engine. For other users, who are interested in browsing the website and learning more about Austria, Austria-Forum has divided its content into several different categories, such as culture, people, scenery, nature and more, with the ultimate goal of keeping users engaged and increasing their session lengths as well as clicks on the website. The link structure of Austria-Forum mostly forms a huge hierarchy. Arriving at the main page users can choose one of $22$ main categories and start navigating the hierarchy downwards to a specific topic (e.g., \emph{main page/nature/fossils/amber}). Overall, nearly $90$ percent of all links within Austria-Forum can be categorized as hierarchical links.

\DatasetHTTPRequestTable
\FigureDatasetDescription

\noindent\textbf{Log Data.} For our analysis we use data that was gathered by logging \textit{HTTP-Requests} on \url{http://www.austria-forum.org}, as well as other domains---such as the outdated \url{http://www.austria-lexikon.at}---which link to it. The observation period of our logs consisted of $59$ days in April, May, and June of $2015$. 

Table \ref{tab:dataset:httprequest} lists the parts of the \textit{HTTP-Requests}, which were logged and provides a typical example \textit{HTTP-Request} of a successful access request to Austria-Forum.

As we are mainly interested in user navigational behavior, we have extensively filtered the logs. First, we filtered the \textit{Content-Type} to only include human-readable \textit{HTML} pages, eliminating \textit{XML}, \textit{templates} and \textit{attachments}.
Second, \textit{Referrers} and \textit{Targets} indicating admin or irregular user behavior, were removed. The removed logs included previewing an edit for a page, pressing the upload button to attach files to articles, or \textit{RSS-Feed-Requests}. Third, we have only kept \textit{Requests} which successfully transmitted a page to the user, indicated by the \textit{Response Code}. Therefore, we have removed all \textit{Requests} with \textit{Response Codes} other than $200$ (OK).

In order to be able to identify pages with multiple \textit{URLs}, \textit{Requests} were normalized by removing the ``\textit{www.}'' prefix as well as trailing slashes ``\textit{/}'' when applicable.
We stripped the data of all entries created by well known \textit{User-Agents} of crawlers, such as GoogleBot, or whenever the \textit{User-Agent} contained a specific substring, such as \textit{crawl}, \textit{slurp}, \textit{spider} or \textit{bot}, which suggested bot activities. Furthermore, to identify bots which do not want to be recognized as such, we removed all entries which had the same \textit{Target} as \textit{Referrer}, which is abnormal behavior as standard page-refreshes usually retains the last \textit{Referrer}. As many bots leave the \textit{Referrer} in their \textit{Requests} empty, all sessions with $4$ clicks and more ($47,312$) that had more than half of its \textit{Referrers} missing were removed. Using this procedure, we removed a little over half ($24,293$) of those sessions. 

The specific method that was used on the server to generate \textit{Session-IDs} is unknown to us. As we assume that the \textit{Remote-IP} as well as cookies are likely considered for generating sessions, it is no simple task to combine, split, recreate and aggregate \textit{HTTP-Requests} into navigational sessions. The number of \textit{Session-IDs}  exceeds the number of \textit{Remote-IPs} by a large margin, which we presume is due to static \textit{IPs} of some users such as schools using the same \textit{IP} for all students, and users with browser add-ons to increase anonymity (so that no \textit{Session-ID} can be mapped to that specific user).
To make sure that sessions by the same user in different periods could be recognized as such, we introduced a time delta which---if exceeded between two requests---indicates the start of a new session. Hence, a smaller delta increases the number of sessions (Figure \ref{fig:dataset:numsessions}). Decreasing delta too far would split sessions at pages where users spent a lot of time, even tough in reality the users were still active in their sessions.

Meiss et al. \cite{meiss2009s} showed that separating \textit{HTTP-Requests} (which they gathered on the entire web) into sessions, can not be done in a clean way solely based on timeouts. Hence, they introduced the concept of logical sessions. In particular, users can have multiple logical session at the same time. For example: browsing domains consisting of mostly images in one tab while navigating on encyclopedias in others. Depending on the domain, average time spent per page varies greatly, as images can be consumed much faster than textual content. In their research they identified a timeout of $15$ minutes as a good approximation of a logical user session.
Since users tend to browse Austria-Forum for research, information, self-improvement, or just to educate themselves further, their sessions can be seen as logical as long as the time between two requests is not exceedingly long. It can be assumed that the time users spend on a page in an encyclopedia can be substantially longer than on an average webpage, due to long (and possibly) complex articles. Taking these factors into consideration, we found that setting our delta to $30$ minutes still split several sessions while granting our users enough time for longer page visits. With delta set to $30$ Minutes, the average session was $1.95$ clicks long (Figure \ref{fig:dataset:avgsession}).

The distribution of sessions can be seen in Figure \ref{fig:dataset:sessiondistr}. It is apparent that the distribution is highly skewed and heterogeneous, indicating many short sessions of few clicks (portrayed by many sessions which are situated low on the $y$-axis) and a few very long sessions (represented by a few sessions in the upper left corner). The short sessions are mostly users who were referred to Austria-Forum by a search engine and either instantly found the information they needed or ceased looking for the needed information on Austria-Forum.

\noindent\textbf{Crawling the Link-Structure.} To compare the navigation behavior of website visitors to the random surfer, we have crawled the whole link structure of Austria-Forum. To this end, we have developed a simple Web crawler that we pointed towards the main page of the website, and which then recursively crawled and followed all encountered (internal) links by pursuing a breadth-first strategy. Some of the encountered links were removed, such as all requests to display the raw Wiki sources for each page that are easily identified by the \textit{skin=raw} parameter in the \textit{URLs}. Further, links to binary files, such as \textit{.mp3}, \textit{.mp4}, \textit{.jpg}, and many more, have been removed as well, as we are only interested in the navigation behavior of users while browsing and exploring the underlying website. 

\noindent\textbf{Limitations.} We were not able to include the clicks of users within the Web Books of Austria-Forum in our study. Further, to simplify the data preprocessing, we cut off active sessions at midnight.

\subsection{Random Surfer}
\label{sec:random_surfer}

\noindent\textbf{Preliminaries.} Mathematically, a random surfer is represented by a random walk on a weighted directed graph. Thus, we start by introducing some basic notion for such random walks.

Let $\bm{A}$ be the weighted adjacency matrix of a directed and weighted graph $G$ with $A_{ij}>0$ if node $j$ points to node $i$ and $0$ otherwise. The value of $A_{ij}$ represents the weight of the link from $j$ to $i$. The weighted out-degree $k_i^+$ of a node $i$ is defined as the sum over the weights of outgoing links:
\begin{equation}
    \label{eq:degree_vec}
    k_i^+ = \sum_{j=1}^{n}A_{ji}.
\end{equation}

Let $\bm{D}$ be a diagonal matrix of weighted out-degrees, so that $d_{ii}=k_i^+$ if $k_i^+ > 0$, otherwise we set $d_{ii}=1$. The matrix $\bm{P}$, defined as
\begin{equation}
\bm{P} = \bm{AD}^{-1},
\end{equation}
\noindent
 is than a transition matrix of a random walk on the weighted directed graph $G$. An element $P_{ij}$ of the matrix defines the probability of a random surfer moving from node $j$ to node $i$.
 
A stationary distribution of a random walk is defined as a probability of finding a random walker at a particular page in the limit of infinitely many steps. Algebraically, the stationary distribution is equal to the right eigenvector corresponding to the largest eigenvalue of the transition matrix $\bm{P}$. If the graph $G$ is strongly connected and the transition matrix does not allow only periodic returns to a given state, then the largest eigenvalue of the matrix $\bm{P}$ is $1$, and the stationary distribution is unique. In the case of a graph $G$ that is not strongly connected, teleportation represents a simple technical solution as it connects each page to every other page with small weight. Teleportation also guarantees that there are not exclusively periodic returns to any given state in the network since there is a constant small probability to remain at the current page after teleporting the surfer to exactly that page. Thus, we therefore include teleportation in our calculations and calculate PageRank vectors of pages from $G$.

The calculation of the PageRank vector of the weighted adjacency matrix simplifies to (details are given in e.g., \cite{newman}):

\begin{equation}
\bm{\pi} = \bm{D}(\bm{D}-\alpha \bm{A})^{-1}\bm{1},
\label{eq:stat}
\end{equation}
where $\alpha \in \left[0,1\right]$ is the damping factor.

\smallskip
\noindent\textbf{Uniform random surfer.} For the uniform random surfer we use the graph $G$, that we crawled from Austria-Forum. We do not set weights to hyperlinks for the uniform random surfer, thus we set $A_{ij}=1$ if node $j$ points to node $i$ and $0$ otherwise.

\smallskip
\noindent\textbf{Pragmatic random surfer.} To create a weighted adjacency matrix containing information of user transitions we first filter out teleportations, meaning transitions which are not present in the adjacency matrix of the network. Afterwards we account for user transitions that we observed in the network adjacency matrix. For that purpose, we apply sublinear scaling to the transition counts, which is a common scaling technique in the field of information retrieval---a word which occurs, for example, $20$ times in an document is not assumed to be $20$ times more significant than a word occurring only once. For navigation we can make an analogous assumption, meaning that $20$ observed transitions from page A to page B does not make this transition $20$ times more significant than a single transition from, for example, page A to page C. In many cases there are several links between any two pages and some of these links are prominently presented in the user interface (e.g., in the navigation bar) inducing bias to the link selection process by users.

Therefore, sublinear scaling seems to be an appropriate approach to account for such situations. We scale the transition counts in the following way. Let $t_{ij}$ be the number of transitions between pages $j$ and $i$. We then calculate scaled transition count $c_{ij}$ as:

\begin{equation}
\label{eq:sublinear_scaling}
c_{i,j} = \begin{cases}
1+\ln{t_{i,j}} &\text{if $t_{i,j} > 0$}\\
0 &\text{otherwise}
\end{cases}
\end{equation}

After scaling down the transition counts we calculate the weighted adjacency matrix for the pragmatic random surfer in the following way. Let $\bm{C}$ be a matrix containing scaled transition counts, with $C_{ij}$ being the scaled number of transitions between pages $j$ and $i$. Further, we define a vector $\bm{v}$ which is a binary vector with $v_i=1$ if the page $i$ has been visited at least once by any of the users. Otherwise we set $v_i=0$. Finally, let $\bm{V}$ be a diagonal matrix with vector $v$ on the diagonal. Then the adjacency matrix of a directed network weighted with the scaled user transition counts can be calculated as follows:

\begin{equation}
\bm{A}=\bm{V}(\bm{A}_u+\bm{C})\bm{V},
\end{equation}
where $\bm{A}_u$ is the adjacency matrix of the unweighted graph as used for the uniform random surfer. After removing all rows and columns consisting of only zeros this results in the adjacency matrix of the induced sub graph, which only includes nodes visited at least once by any user and all edges between those nodes (independent if traversed by any user or not). Now, the stationary distribution $\bm{\pi}$ may be calculated as given by Equation~\ref{eq:stat}.

\smallskip
\noindent\textbf{Lateral random surfer.} 
We represent the lateral random surfer only through its stationary distribution. The stationary distribution of the lateral random surfer we calculate by simply normalizing page views we directly obtained from the server access logs. Specifically, we do not have a random surfer in this case, but observe the resulting stationary distribution of an underlying random navigation process.

\subsection{Gini coefficient}
\label{sec:meth_gini}
The Gini coefficient is a metric for measuring inequality of a distribution. It computes the area between the Lorenz curve~\cite{lorenzcurve} and the uniform distribution. Higher values indicate a larger difference and higher inequality. For our analyses, we calculate the Gini coefficient for the stationary distributions of all three random surfer types.

\section{Results \& Discussion}
\label{sec:results}

\FigureHeatmap
\FigureBiasFactors

In our experiments we are interested in comparing and analyzing the differences and commonalities between the uniform random surfer model, the pragmatic random surfer model and the lateral random surfer model (cf. Section~\ref{sec:random_surfer}). We use the power iteration method~\cite{brin} to calculated the PageRank vector. In the first experiments we set $\alpha$ to a fixed value of $0.85$. This correspond to teleportation probability of $15\%$, analogously to the original PageRank algorithm~\cite{brin}. Hence, the damping factor corresponds to the probability of a user to keep navigating over adjacent pages at each step. In later experiments we analyze the influence of various values for $\alpha$.
Figure~\ref{fig:heat} depicts the different correlations between the stationary distributions of all three random surfer models. In particular, the Pearson correlation coefficient between the uniform and pragmatic random surfer of $\rho=$\corUnifPrag indicates nearly perfect positive correlation. Thus, this correlation analysis shows that there is a considerable overlap between the behaviors of the uniform and pragmatic random surfer models. In conclusion, the uniform random surfer model appears to be a very good approximation of the pragmatic random surfer---which in our case represents a proxy for user behavior---on Austria-Forum.

On the other hand, the uniform ($\rho=$\corUnifLat) and pragmatic ($\rho=$\corPragLat) random surfer models exhibit only weak levels of correlation to the lateral random surfer.
Further, the heat maps depicted in Figure~\ref{fig:heat} strengthen our findings, as the lateral random surfer, representing users entering the website from for instance search engines, exhibits higher probabilities to visit pages which are rated as unimportant by the uniform or the pragmatic random surfer.
In other words, they are pointed directly to specific pages without the need to navigate the hierarchy of the website. Thus, search engines appear to reduce the need for users to navigate (hierarchical) website structures and therefore are an important factor to include in (future) analyses of user navigation behavior.

\vspace{1ex} \noindent \framebox{\parbox{\columnwidth}{ \textbf{Finding 1:} Uniform random surfer is a very good model of user navigational behavior in our dataset. It correlates almost perfectly with the pragmatic random surfer constructed from the clickstream data. On the other hand, both uniform and pragmatic random surfer significantly differ from the lateral random surfer, which also reflects user visits from search engines.}} \vspace{1ex}

In further experiments we varied $\alpha$ (damping factor of PageRank) and found that with lower values of $\alpha$ (e.g, $\alpha=0.2$) the correlation between uniform and lateral random surfer increases from  $\rho=$\corUnifLat to $\rho=0.49$, which suggests that higher teleportation probabilities better capture the lateral user access from search engines. However, at the same time the correlation between the pragmatic and the lateral random surfer decreases from $\rho=$\corPragLat to $\rho=0.29$ for $\alpha=0.2$ while the correlation between the uniform and the pragmatic remains stable and above $0.9$. This result suggests that the lateral access to a website can not be solely captured by a random surfer with teleportation. Rather we need to extend this basic model. For example, we could use the basic model to also model navigational sessions. In this model teleportation probability increases with every new click to account for an increased likelihood of switching to a new session as the user makes progress in the current session.

\vspace{1ex} \noindent \framebox{\parbox{\columnwidth}{ \textbf{Finding 2:} To capture the lateral access to a website from a search engine we need a new kind of random surfer model.}} \vspace{1ex}

Furthermore, we calculated and compared the ratios of stationary probabilities for each page and between all combination of three random surfer models to investigate commonalities and differences between them (see Figure~\ref{fig:biasfac}). Although the uniform and pragmatic random surfer models exhibit a Pearson correlation coefficient of almost $\rho=1$, there are a few pages with a ratio of $10$ or $0.1$. This means that those pages are $10$ times more (less) important for the pragmatic random surfer than for the uniform random surfer. Figure~\ref{fig:biasfac:unif_prag} depicts a specific trend showing that pages with a low value in the stationary distribution of the uniform random surfer often obtain much higher values with the pragmatic random surfer. This difference is compensated by somewhat smaller importance for the pragmatic random surfer of the mid and high important pages for the uniform random surfer.

When comparing the ratios of the uniform and lateral random surfer models, we can see even stronger tendencies than in our previous analysis. The general shape of the differences remains the same, meaning less important pages for the uniform random surfer become more important for the lateral one, but the magnitude of the differences is larger now and goes in some cases up to $100$. Similar observation can be made for the most important pages for the uniform random surfer, which now become less important also in some cases by a factor of $100$ (see Figure~\ref{fig:biasfac:unif_lat}). Finally, Figure~\ref{fig:biasfac:prag_lat} depicts the ratios of the pragmatic random surfer compared to the lateral random surfer. Again, we make a very similar observation as in the case of differences between the uniform and the lateral random surfer.

\vspace{1ex} \noindent \framebox{\parbox{\columnwidth}{ \textbf{Finding 3:} Although the assessment of individual page importance between the uniform random surfer and the pragmatic random surfer differs in some cases by a factor of $10$, the assessments are generally very well aligned. The differences in assessments between the uniform and the pragmatic on the one side, and the lateral random surfer on the other side are often very large (factor of $100$). The general alignment in the assessment between the lateral and other two models is not given in our dataset.}} \vspace{1ex}

The Lorenz curves of the stationary distribution of all three random surfers are shown in Figure~\ref{fig:gini}. The uniform random surfer achieves a Gini coefficient of~\giniUnif. With a value of~\giniPrag, the pragmatic random resulted in a lower coefficient. This means that the inequality in the stationary distribution of the pragmatic random surfer is lower than that of the uniform random surfer. In other words, the imbalances in the individual page importance are reduced as low importance page become more important, and vice versa highly important pages are less important for the pragmatic random surfer. Finally, the lateral random surfer exhibits the comparatively lowest Gini coefficient of \giniLat. Due to the bias towards more specific pages located in lower levels of the website hierarchy in the lateral random surfer, this type of the random surfer is less likely to be directed towards highly popular pages as compared to the uniform random surfer. 

\FigureGini

\vspace{1ex} \noindent \framebox{\parbox{\columnwidth}{ \textbf{Finding 4:} The imbalances in the relative page importances are reduced for the pragmatic random surfer (only slightly) and for the lateral random surfer (significantly) as compared to the uniform random surfer. Direct lateral access from search engines towards more specific pages reduces the degree to which a random surfer is directed towards high importance pages.}} \vspace{1ex}

\section{Conclusions \& Future Work}
\label{sec:conclusions}

In this paper we presented new insights into the commonalities and differences between a uniform random surfer, a user clickstream biased (pragmatic) random surfer and a page visits biased (lateral) random surfer. 
We compared the navigation behavior of these three different random surfer models in an online encyclopedia, namely Austria-Forum. Using empirical user data we showed that the random surfer represents a good approximation of navigational user behavior for the investigated website---allowing researches to conduct user navigation experiments using a simple random surfer without the need to collect user clickstreams. 
Due to the low correlation between uniform and lateral random surfer we conclude, that the hierarchical structure of a website does not play such an important in role in terms of user navigation as it did before the rise of search engines. The majority of users enter the website using a search engine and leave after consuming the landing page. Hence, the uniform random surfer model is a good approximation of user navigation as long as no search engines are involved. However, hierarchical structures are needed for most search engines to rank the results of search queries. Nevertheless, the observed behavior leads to the question if website administrators should additionally provide page recommendations to keep users navigating their page. 

Further experiments with varying teleportation probabilities (i.e., lower $\alpha$) for the random surfer show that we can increase the correlation of stationary distributions between the uniform and lateral random surfer, but at the same time decrease the correlation between the pragmatic and the lateral random surfer. These differences in modeling navigational user behavior with and without search engines  represent the directions for future work for modeling and hence optimizing navigational potential of a website.

Our results represent important insights for website administrators, search engine providers and researchers who want to broaden their understanding of user navigation and the models thereof. 
The contributions of this paper may serve as an interesting input to modify the models and for example link recommendation algorithms to influence navigational behavior of users. With this work we contribute to the analysis of user navigational behavior by (i) providing a comparison of random surfer model data with clickstream data, (ii) a thorough analysis of the differences between these random surfer models on a Web Encyclopedia and (iii) presenting a methodology that allows us to estimate the optimization potential of a website in terms of keeping users navigating on the website as long as possible.

\noindent\textbf{Future Work.} In future, we plan to verify our results on other websites where user clickstreams are available (e.g., the English Wikipedia). Furthermore, we want to use our model to test different types of biases introduced into the front end (e.g., recommendations of other pages) of a website to analyze to which extent such biases are able to influence users in their navigation. Another idea is to modify the order of recommendations in a recommendation network and analyze---based on the assumption that recommendations on the top are clicked more often~\cite{positionbias}---the influence thereof.

\balance

\section{*Acknowledgments}
This research was in part funded by the FWF Austrian Science Fund research project "Navigability of Decentralized Information Networks" (P 24866). We thank Gerhard Wurzinger for providing access to Austria-Forum server logs.
\vspace{-0.5em}
\balance
\bibliographystyle{abbrv}

\end{document}